\documentclass[onecolumn]{revtex4-2}
\usepackage{graphicx}
\usepackage{graphics}
\usepackage{tabularx}
\usepackage{gensymb}
\usepackage{amsmath}
\usepackage{longtable}
\usepackage{multirow}
\usepackage{array}

\usepackage{url}
\usepackage{color}
\usepackage[usenames,dvipsnames]{xcolor}
\usepackage[colorlinks,hyperindex]{hyperref}
\hypersetup{colorlinks,citecolor=blue,linkcolor=blue,urlcolor=blue}

\usepackage{threeparttable}
\begin{document}

\title{Electronic processes in collisions between nitrogen ions and hydrogen atoms}

\author{C. C. Jia$^{2,3}$}
\author{Y. Y. Qi$^{4}$}
\author{J. J. Niu$^{1}$}
\author{Y. Wu$^{2}$}
\author{J. G. Wang$^{2}$}
\author{A. Dubois$^{3}$}
\author{N. Sisourat$^{3}$}
\email{nicolas.sisourat@sorbonne-universite.fr}
\author{J. W. Gao$^{1}$}
\email{gaojunwen@hznu.edu.cn}

\affiliation{\rm{$^{1}$School of Physics, Hangzhou Normal University, Hangzhou 311121, China}}

\affiliation{\mbox{\rm{$^{2}$}Key Laboratory of Computational Physics, Institute of Applied Physics and Computational Mathematics, 100088, Beijing, China }}

\affiliation{\mbox{\rm{$^{3}$}Sorbonne Universit$\acute{e}$, CNRS, UMR 7614, Laboratoire de Chimie Physique-Mati$\grave{e}$re et Rayonnement, 75005, Paris,  France}}

\affiliation{\mbox{\rm{$^{4}$}College of Data Science, Jiaxing University, Jiaxing 314001, China}}

\date{\today}

\begin{abstract}

In order to interpret and predict the behavior and properties of fusion plasma, accurate cross sections for electronic processes in collisions between plasma impurities and atomic hydrogen are required. In this work, we investigate the electron capture (or charge exchange), target excitation, and ionization processes occurring in collision of ${\rm N}^{4+}$ with atomic hydrogen in a broad energy domain ranging from 0.06 to 225 keV/u. We consider ${\rm N}^{4+}$ ground state ${\rm N}^{4+} (2s)$ and also ${\rm N}^{4+} (2p)$ since the impurities in the edge plasma environment may be excited due to collisions with electrons and ions/atoms. Total and partial cross sections in both spin-averaged and spin-resolved cases are calculated using a two-active-electron semiclassical asymptotic-state close-coupling approach. For electron capture cross sections the present results show the best overall agreement with available experimental data for both total and partial cross sections, and the origins of observed discrepancies are discussed. Furthermore, we provide new data for target excitation and ionization processes, which are essential to improve our understanding of this relevant collision system. %The International Atomic Energy Agency (IAEA) has recently published a report highlighting the importance and the scarcity of such data. Our work therefore will allow a better modeling and thus understanding of magnetically confined fusion plasma. 

\end{abstract}

\maketitle

\section{Introduction}

In magnetic confinement fusion reactors, the injection of a neutral beam, typically hydrogen or deuterium, is a standard method to either heat or diagnose the plasmas~\cite{anderson2000, delabie2010}. The latter employs emissions, so called Charge Exchange Recombination Spectroscopy \cite{Isler_PRL1977, Isler_1994}, from plasma impurities after charge transfer with a neutral beam particle. It is routinely used for the measurement of ion temperature, plasma rotation and impurity densities in tokamak plasmas~\cite {Isler_1994, Hellermann_2005, McDermott_2018}. On the other hand, impurity ions are deliberately seeded into fusion plasmas for a variety of purposes, and it is increasingly utilized, as large-scale experimental devices like International Thermonuclear Experimental Reactor (ITER) emerge, to redistribute power from the plasma core to the reactor wall, reducing heat stress and potential damage to plasma-facing components \cite{hill2022summary,Bonanomi_2018,Komm_2019,Tan_pop,Kallenbach_2010}. To avoid long time fuel storage by co-deposition, the use of nitrogen, argon, neon seeding are excellent choices \cite{Kallenbach_2010}, because it behaves as a weak-recycling, non-intrinsic impurity \cite{McCracken1997} and its spectroscopic emission can be observed by a wide range of standard diagnostic tools~\cite{LaBombard_2017}. In order to interpret and predict the behavior and properties of impurities in fusion plasma, a significant amount of modelling, involving the evaluation of accurate cross sections for electron capture, target excitation, and ionization of impurities in collisions with atomic hydrogen is required~\cite{Marchuk_2014, Hill_2023}.

In the present work, we investigate collisions between N$^{4+}$ ion in its ground and first excited states and atomic hydrogen leading to electron capture,
\begin{eqnarray}
	{\rm N}^{4+} (2s/2p) + {\rm H}(1s) \to {\rm N}^{3+}(n\ell n'\ell'~^{1, 3}L)(n, n'\ge 2 ) +  {\rm H}^{+},
	\label{eqa}
\end{eqnarray}
target excitation,
\begin{eqnarray}
	{\rm N}^{4+} (2s/2p) + {\rm H}(1s) \to {\rm N}^{4+}(n\ell)(n\ge 2 ) +  {\rm H}^{*}(n\ell'),
	\label{eqb}
\end{eqnarray}
and ionization,
\begin{eqnarray}
	{\rm N}^{4+} (2s/2p) + {\rm H}(1s) \to {\rm N}^{4+}(n\ell)(n\ge 2 ) +  {\rm H}^{+} + e^{-}.
	\label{eqc}
\end{eqnarray}
Since the impurities in the edge plasma environment may be excited due to collisions with electrons and ions/atoms \cite{aumayr1991ev}, the study of excited N$^{4+}(2p)$ in collisions with atomic hydrogen is also of relevance. However, to our knowledge, no detailed experimental or theoretical investigation related to excited N$^{4+}(2p)$ projectile has been reported so far. For ground state N$^{4+}(2s)$ ion and atomic hydrogen collisions, numerous experimental and theoretical studies have been conducted over the past decade. Total SEC cross sections were measured by Seim \textit{et al.} \cite{Seim_1981} in the energy range of 1.1 - 3.6 keV/u, by Crandall \textit{et al.} \cite{Crandall_1979} from 1 to 7 keV/u, and by Phaneuf \textit{et al.} \cite{Phaneuf_1978} from 40 to 120 keV/u. In these early experiments, the cross sections were obtained by passing a beam of N$^{4+}$ ion through a high-temperature oven of dissociated hydrogen. Furthermore, using the merged-beam technique, Huq \textit{et al.} \cite{Huq_1989} and Folkerts \textit{et al.} \cite{Folkerts_1995} have measured total SEC cross sections for the low energy range of 1 - 982 eV/u. Nonetheless, there is still a lack of experimental data for energies ranging from 10 to 40 keV/u. For the state-selective SEC cross sections, investigations are more scarce. McCullough \textit{et al.} \cite{McCullough_1995} conducted the first measurement using translational energy spectroscopy (TES) within the energy range of 0.25 - 1.6 keV/u. Bliek \textit{et al.} \cite{Bliek_1998} provided the initial insights on spin-resolved state-selective cross sections in the energy range of 1 - 4 keV/u by means of photon emission spectroscopy. For ionization of atomic hydrogen by N$^{4+}$ collisions, the only available set of experimental data was reported by Shah and Gilbody \cite{Shah_1981} using a crossed-beam technique for energies from 30 to 165 keV/u.

On the theoretical side, using molecular orbital (MO) approaches, the SEC cross sections for N$^{4+}(2s)$ and H collisions were reported by a number of authors \cite{Feickert_1984_APJ, Shimakura_1992, Folkerts_1995, Stancil_1997, Zygelman_1997_PRA} at various levels of sophistication, for low impact energies ranging from 10$^{-6}$ to 10 keV/u. Sidky \textit{et al.} \cite{Sidky_1999} and Cabrera-Trujillo \textit{et al.} \cite{Cabrera-Trujillo_2002_PRA} extended the cross section calculations to intermediate energies up to 25 keV/u using the atomic orbital close-coupling (AOCC) method and electron-nuclear dynamics (END) formalism, respectively. For high impact energies, calculations using the classical trajectory Monte Carlo (CTMC) method \cite{Olson_1977, Purkait_2001_JPB} and the boundary corrected continuum intermediate state approximation \cite{Purkait_1999_PRA, Sounda2006} were reported in the energy range of 40 - 200 keV/u. Despite the intensive investigation, considerable discrepancies still exist between available experimental and theoretical results. Moreover, no experimental data and theoretical predictions for target excitation exist. This lack of data impedes the modeling of fusion plasma.

%a two-active-electron close-coupling approach
In the present work, the electron capture, target excitation, and ionization processes (Eqs.(\ref{eqa}-\ref{eqc})) are investigated in a wide energy domain ranging from 0.06 to 225 keV/u. This energy range covers the typical energies of many astrophysical and laboratory plasmas, including the radiation losses and neutral beam heating efficiencies in tokamak plasmas. We use a two-active-electron semiclassical asymptotic-state close-coupling (SCASCC) approach with a configuration interaction treatment to deal with electronic correlation. The total and $n\ell$-resolved cross sections in both spin-averaged and spin-resolved cases are presented and compared with previous experimental measurements and theoretical calculations where available. Possible reasons for the disagreements with existing data are also discussed. We therefore provide a full set of consistent data, which can be employed in various plasma modeling.

The present paper is organized as follows. In the next section we briefly outline the SCASCC method used in the present calculations. In Sec.~\ref{sec3}, we present a detailed analysis of electron capture, target excitation, and ionization cross sections, including direct comparisons with available experimental and theoretical results. A brief conclusion follows in Sec.~\ref{sec4}. Atomic units are used throughout, unless explicitly indicated otherwise.

\section{Theory}

In the present work, we have calculated electron capture, target excitation and ionization cross sections in N$^{4+}$ and H collisions using a two-active-electron SCASCC approach, which was previously described, see e.g. in~\cite{nicolas_2011,Gao_2017,Gao_2023_Apj,Gao_PRA_2024}. Here we outline only briefly the main features of the approach. The two-electron time-dependent Schr$\rm\ddot{o}$dinger equation (TDSE) is written as
\begin{equation}\label{eq2}
\small
\left[ { H-i\frac { \partial  }{ \partial t }  } \right] \Psi(\vec{r}_1,\vec{{ r }}_2,t)=0,
\end{equation}
with the electronic Hamiltonian
\begin{equation}\label{eq3}
\small
\begin{split}
H&=\sum_{i=1,2}\left[-\frac{1}{2}\nabla^2_i+V_T(r_i)+V_P({r}_{i}^{p})\right]+\frac{1}{\left|\vec{ { r }_{ 1 } }-\vec{ { r }_{ 2 } }\right|},
\end{split}
\end{equation}
where $\vec{ { r }_{ i } }$ and $\vec{r_{i}}^{p}=\vec{ { r }_{ i } }-\vec{R}(t)$ are the position vectors of the electrons with respect to the target and the projectile, respectively. The relative projectile-target position $ \vec{R}(t)$ defines the trajectory,  with $\vec{R}(t)=\vec{b}+\vec{v}t$ in the usual straight-line, constant velocity approximation where
$\vec{b}$ and $\vec{v}$ are the impact parameter and relative velocity vector, respectively. The term $V_{T}$ ($V_{P}$) is the electron-target (-projectile) nucleus (and inner electrons, in the frozen core approximation) potential,
\begin{equation}\label{eq4}
\begin{split}
V_{T}(r_{i})=-\frac{1}{r_i},\hspace{5pt} V_{P}({r}_{i}^{p})=-\frac{5}{r_i^{p}}-\frac{2}{r_i^{p}}(1+\alpha r_i^{p})e^{-\beta r_i^{p}},
\end{split}
\end{equation}
with the variational parameters $\alpha = 4.3982$ and $\beta = 8.7964$ are taken from~\cite{Errea_2002,Zhang_2020_PRA}.

The Schr$\rm\ddot{o}$dinger equation is solved by expanding the wavefunction onto a basis set composed of states of the isolated collision partners (i.e., asymptotic states),
\begin{equation}\label{eq5}
\begin{split}
\Psi (\vec{r_{1}}, \vec{r_{2}}, t)=&\sum_{k=1}^{N_{T}} \sum_{l=1}^{N_{P}} a_{kl}^{TP}(t)[ \phi_{k}^{T}(\vec{r_{1}}) \phi_{l}^{P}(\vec{r_{2}}, t ) \pm \phi_{k}^{T}(\vec{r_{2}}) \phi_{l}^{P}(\vec{r_{1}}, t ) ]   e^{-i(E_{k}^{T}+E_{l}^{P})t}\\
&+\sum_{j=1}^{N_{PP}}a_{j}^{PP}(t)\Phi_{j}^{PP}(\vec{r_{1}}, \vec{r_{2}}, t)e^{-iE_{j}^{PP}t},
\end{split}
\end{equation}
where $T$ ($P$) and $PP$ superscripts denote states and corresponding energies for which one or two electrons are on the target (projectile), respectively.
Note that, in the present work, the two-electron states of target (H$^{-}$) are not included in the calculations, since the formation of H$^{-}$ is 
expected to be negligible compared to the processes in Eqs.~(\ref{eqa}-\ref{eqc}).
The $\pm$ sign in the last part of Eq.~(\ref{eq5}) stands for the singlet and triplet spin states, respectively, and the wave functions $\Phi_{j}^{PP}$ are related to the corresponding spin symmetry. For both electrons, the projectile states contain plane-wave electron translation factors (ETFs),
$e^{i\vec{v}\cdotp\vec{r_{i}}- i\frac{1}{2}v^{2}t}$, ensuring Galilean invariance of the results.
The insertion of Eq.~(\ref{eq5}) into (\ref{eq2}) results in a system of first-order coupled differential equations, which can be written in matrix form as
\begin{equation}\label{eq6}
i \frac{d}{dt}\textbf{a}(t) =\textbf{S}^{-1}(\vec{b}, \vec{v}, t)\textbf{M}(\vec{b}, \vec{v}, t)\textbf{a}(t),
\end{equation}
where $\textbf{a}(t)$ is the column vector of the time-dependent expansion
coefficients, i.e. $a^{PP}$ and $a^{TP}$ in Eq.~(\ref{eq5}) and \textbf{S}, \textbf{M} are the overlap and coupling matrices, respectively. We should emphasize here that we include explicitly in \textbf{M} all bielectronic couplings, and notably the complex two-center ones which include the ETF for both electrons. We use the same strategy, i.e. no neglect of any couplings, to obtain  the two-electron (PP) states in the diagonalisation stage, going well beyond the Hartree-Fock approach. Our approach represents therefore a configuration interaction (CI) approach.

The coupled equations Eq.~(\ref{eq6}) are solved using the predictor-corrector variable-time-step Adams-Bashford-Moulton method for a set of initial conditions: initial state $i$, and given values of $b$ and $v$. The state-to-state integral cross sections for a transition are calculated as
\begin{equation}\label{eq8}
	\sigma_{fi}(v)=2\pi \int_{0}^{+\infty}b\left|a_{f}(+\infty)\right|^{2}db.
\end{equation}

In our approach in which total spin is conserved, the coupled equations Eq.~(\ref{eq6}) are solved independently for singlet and triplet symmetries corresponding to the two possible total spin states of the collision system. The corresponding cross sections are afterwards multiplied by their spin statistical weights ($\frac{1}{4} $ for the singlet states and $\frac{3}{4} $ for the triplet ones) to allow for a direct comparison with the experimental data. The spin-averaged cross sections are obtained as the sum of the singlet and triplet results according to their statistical weights.

\section{\label{sec3}Results and discussions}

To carry out the computations of cross sections for N$^{4+}$-H collisions, we first construct the atomic states of H, N$^{4+}$ and N$^{3+}$, to be included in the expansion Eq.~(\ref{eq5}). We express these states in terms of Gaussian-type orbitals (GTOs),
\begin{equation}\label{eq_gto}
G_{\alpha,\ell,m}(\vec{r})=r^\ell e^{-\alpha r^2} Y_{\ell,m}(\theta, \phi),
\end{equation}
for the one-electron ($\phi^{T}$, $\phi^{P}$ in Eq.~(\ref{eq5}) ) states and products of the same GTOs for the two-electron ($\Phi^{PP}$) ones. The number of GTOs and their  exponents $\alpha$ are optimized to get correct binding energies of the relevant states populated during the collision.

In the present calculations, a set of 56 GTOs (12 for $\ell$ = 0, 8 $\times$ 3 for $\ell$ = 1, and 4 $\times$ 5 for $\ell$ = 2) is used on the N$^{5+}$ center, and 27 GTOs (9 for $\ell$ = 0 and 6 $\times$ 3 for $\ell$ = 1) are located on the H center. The basis sets allow the inclusion of 1927 states and pseudo-states (855 (H, N$^{4+}$) and 1072 (N$^{3+}$) states) in the calculations for spin singlet symmetry, and 1919 (855 (H, N$^{4+}$) and 1064 (N$^{3+}$) states) for triplet symmetry. These states can describe elastic, single electron capture, excitation, as well as ionization through the inclusion of pseudostates with energy lying above ionization thresholds.

We have checked our results presented in the following by repeating the calculations with a larger basis set on the N$^{5+}$ center built from 64 GTOs (14 for $l$ = 0, 10 $\times$ 3 for $l$ = 1, and 4 $\times$ 5 for $l$ = 2) and on the H center built from 32 GTOs (11 for $l$ = 0 and 7 $\times$ 3 for $l$ = 1). For both N$^{4+}(2s)$ and N$^{4+}(2p)$ projectiles, similar convergences of cross sections were obtained. Total and dominant $n\ell$-selective (N$^{3+}(2s3\ell$ $^{1}\rm{L})$ and N$^{3+}(2p3\ell$ $^{1,3}\rm{L})$) SEC cross sections from these two sets agree with each other within 2\% at low and intermediate energies, and about 10\% for energies above 100 keV/u. For the cross sections reported in Figs.~\ref{fig6} and \ref{fig7}, the convergence is better than 10\% for the target excitation and about 5\% for the ionization at intermediate energies. In the high energy region, the convergence is evaluated to be about 3\% for the dominant channel H ($2p$), and about 20\% for the other relative weak excitation channels and ionization processes. For energies below 1 keV/u, the convergence for both target excitation and ionization is found to be about 80\% at worst where, however, the cross sections are negligibly small compared to their counterparts at higher energies, see Figs.~\ref{fig6} and \ref{fig7}.

\subsection{Spin-averaged total and $n\ell$-selective electron capture cross sections}

\begin{figure}[!t]
\begin{center}
\includegraphics[scale=0.5]{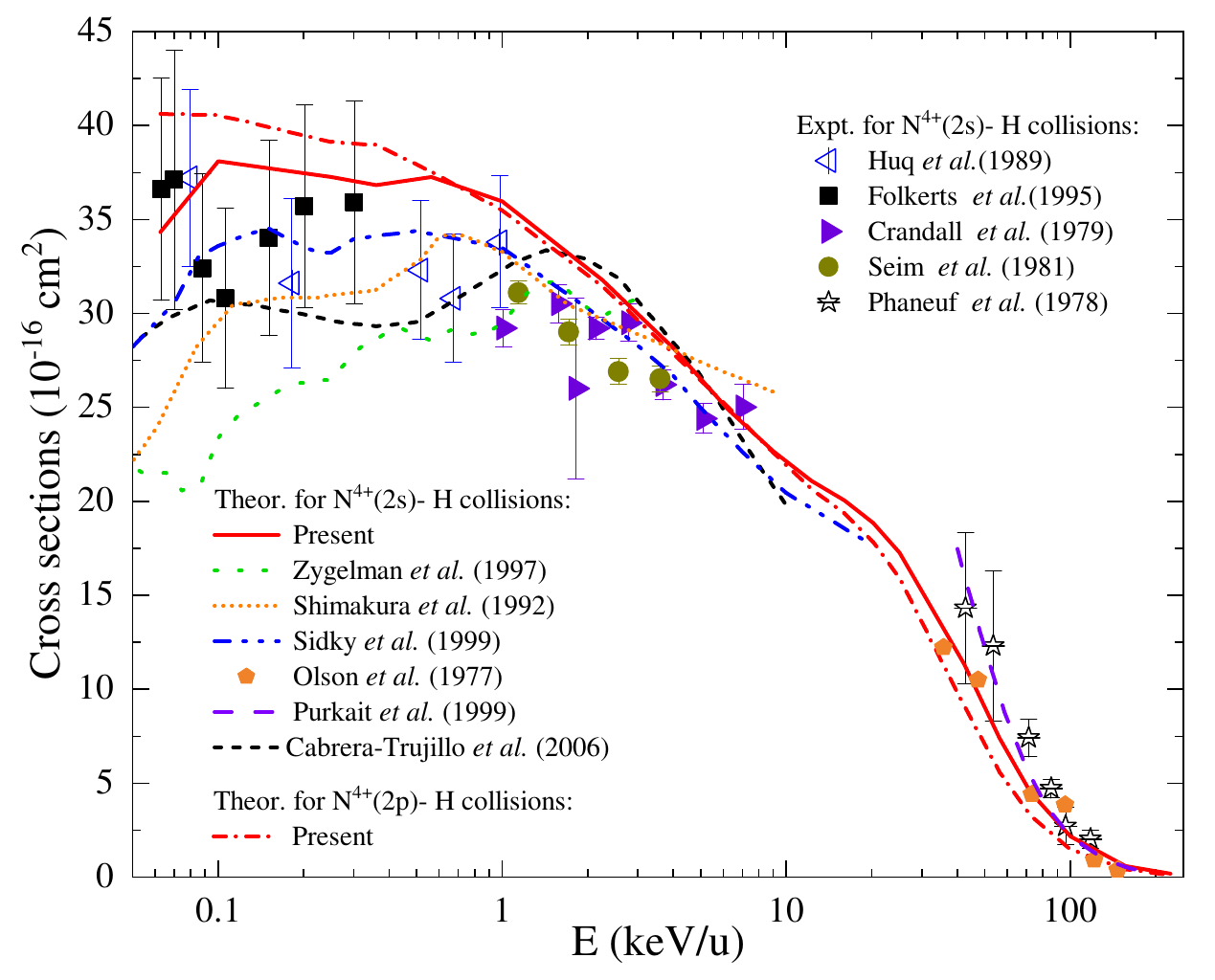}\\
\caption{Spin-averaged total electron capture cross sections of ${\rm N}^{4+} (2s)$ and H collisions as a function of impact energy. The theoretical results are from the present calculations (red solid line), MOCC calculations by Zygelman \textit{et al.}~\cite{Zygelman_1997_PRA} (green dotted line) and Shimakura \textit{et al.}~\cite{Shimakura_1992} (orange short-dotted line), AOCC calculations of Sidky \textit{et al.}~\cite{Sidky_1999} (blue dash-dot-dotted line), CTMC calculations of of Olson \textit{et al.}~\cite{Olson_1977} (yellow solid pentagons), calculations of the boundary corrected continuum intermediate state approximation from Purkait \textit{et al.}~\cite{Purkait_1999_PRA} (purple dashed line), and END calculations of Cabrera-Trujillo \textit{et al.}~\cite{Cabrera-Trujillo_2002_PRA} (black short-dashed line). The experimental results are from Huq \textit{et al.}~\cite{Huq_1989} (blue open leftward triangles), Folkerts \textit{et al.}~\cite{Folkerts_1995} (black solid squares), Crandall \textit{et al.}~\cite{Crandall_1979} (purple solid rightward triangles), Seim \textit{et al.}~\cite{Seim_1981} (dark-yellow solid circles) and Phaneuf \textit{et al.}~\cite{Phaneuf_1978} (black open stars). The present spin-averaged total electron capture cross sections of ${\rm N}^{4+} (2p)$ and H collisions are also presented by red dash-dotted lines.}\label{fig1}
\end{center}
\end{figure}

\begin{figure}[!h]
\begin{center}
\includegraphics[scale=0.5]{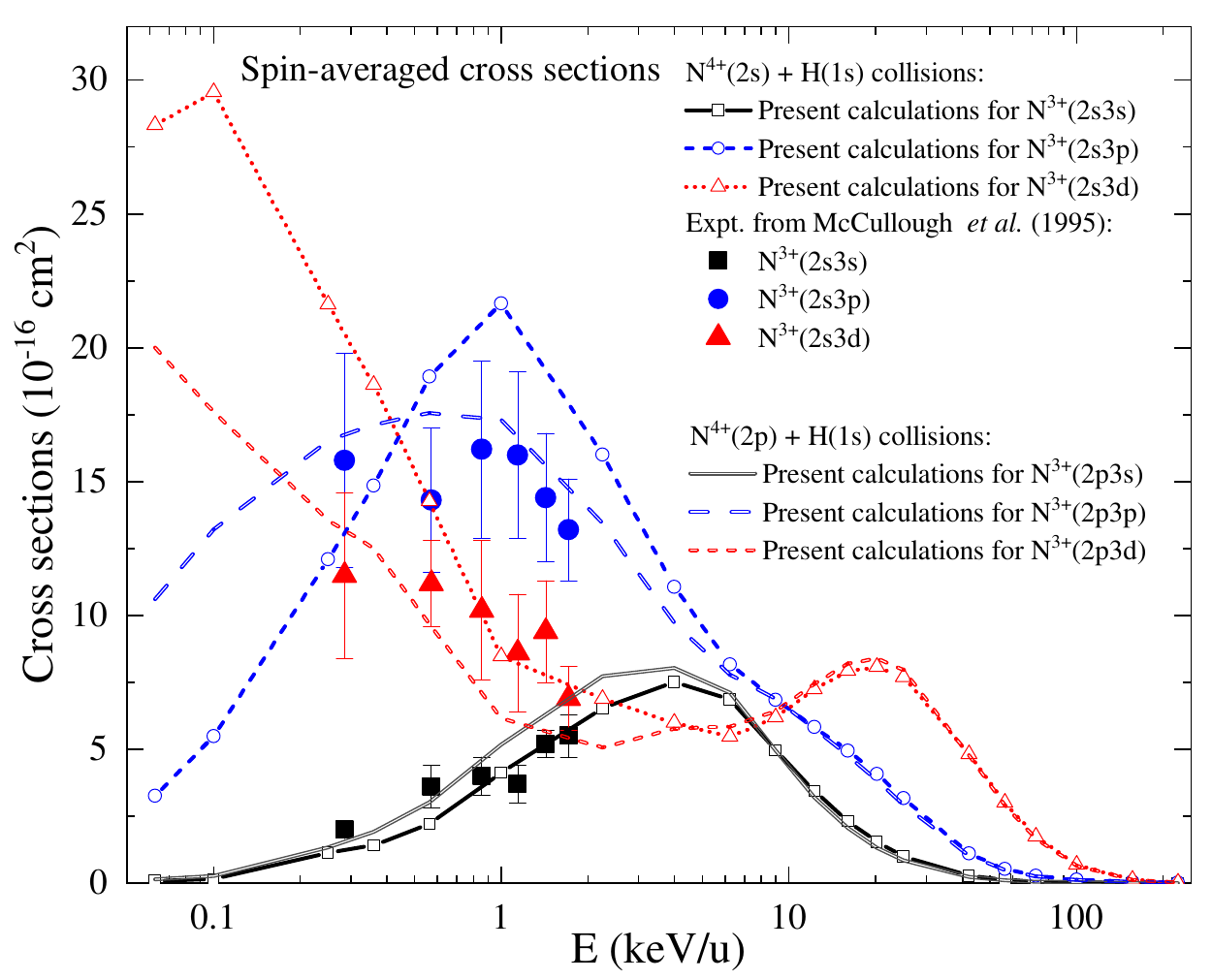}\\
\caption{Spin-averaged $3\ell$-selective electron capture cross sections. For N$^{4+}(2s)$ + H($1s$) collisions, the present calculations are plotted as lines with symbols; the experimental results of McCullough \textit{et al.}~\cite{McCullough_1995} are shown as symbols and error bars. The present results for ${\rm N}^{4+} (2p)$ and H collisions are presented by lines.}\label{fig2}
\end{center}
\end{figure}

In Fig.~\ref{fig1}, we present our calculated spin-averaged total electron capture cross sections as a function of impact energy in the range 0.06 – 225 keV/u, together with previous experimental~\cite{Huq_1989,Folkerts_1995,Crandall_1979,Phaneuf_1978}, and theoretical~\cite{Zygelman_1997_PRA,Shimakura_1992,Sidky_1999,Olson_1977,Purkait_1999_PRA,Cabrera-Trujillo_2002_PRA} results for comparison. Our cross sections show a weak maximum around 0.1 keV/u, and decrease for increasing energies. The presence of a shoulder at about 20 keV/u is due to the increasing importance of the population of high-$\ell$ states, i.e. N$^{3+}(2s3d)$, at high impact energies, as discussed below (see Fig.~\ref{fig2}). It can be observed that our results are comparable to the measurements of Folkerts \textit{et al.}~\cite{Folkerts_1995}, Huq \textit{et al.}~\cite{Huq_1989}, Crandall \textit{et al.}~\cite{Crandall_1979} and Phaneuf \textit{et al.}~\cite{Phaneuf_1978} in the respective overlapping energy regions. Overall, there is a general agreement within the uncertainties of the experiments. However, our cross sections are slightly higher than the experimental data of Seim \textit{et al.}~\cite{Seim_1981} and Crandall \textit{et al.}~\cite{Crandall_1979} for energies ranging from 1 to 4 keV/u. Moreover, other small discrepancies are seen. For instance, Folkerts \textit{et al.}~\cite{Folkerts_1995} highlighted an apparent dip in their measured total cross section near 0.1 keV/u, a feature that has not been seen in any theoretical calculations~\cite{Zygelman_1997_PRA,Shimakura_1992,Sidky_1999,Olson_1977,Purkait_1999_PRA,Cabrera-Trujillo_2002_PRA}, including our own. Comparing with available theoretical results, our results are in good agreement with the CTMC calculations of Olson \textit{et al.}~\cite{Olson_1977} for energies higher than 30 keV/u, with results from END formalism of Cabrera-Trujillo \textit{et al.}~\cite{Cabrera-Trujillo_2002_PRA} for energies ranging from 2 to 7 keV/u. Below 2 keV/u, our total cross sections are larger than the END results, as well as the AOCC calculations of Sidky \textit{et al.}~\cite{Sidky_1999} and the MOCC calculations of Shimakura \textit{et al.}~\cite{Shimakura_1992}. The latter deviation is probably due to insufficient channels included in these previous calculations~\cite{Shimakura_1992,Sidky_1999,Cabrera-Trujillo_2002_PRA}. Moreover, the MOCC calculations by Zygelman \textit{et al.}~\cite{Zygelman_1997_PRA} were intended primarily for lower energies where rotational coupling was not included, which results in a further underestimation of the cross sections. Note finally that according to~\cite{Sidky_1999} the trajectory effect in the AOCC calculations makes about 10\% difference of the cross sections at 0.07 keV/u, and this effect becomes negligible as impact energy increases to 0.3 keV/u and higher energies.

The present spin-averaged total electron capture cross sections for N$^{4+}(2p)$ + H($1s$) collisions are also displayed in Fig.~\ref{fig1}. The results are very close to those of N$^{4+}(2s)$ + H($1s$) collisions, except for energies below 0.5 keV/u. In the latter energy region, the cross sections for N$^{4+}(2p)$ + H($1s$) collisions exceed those of N$^{4+}(2s)$ + H($1s$) by up to 15\%.

In Fig.~\ref{fig2}, we present the spin-averaged $3\ell$-selective electron capture cross sections for N$^{4+}(2s)$ + H($1s$) collisions, compared with the experimental data of McCullough \textit{et al.}~\cite{McCullough_1995}. Our results show that the cross sections for electron capture to N$^{3+}(2s3d)$ dominate for energies below 0.4 keV/u and above 10 keV/u. In between, the results for electron capture to N$^{3+}(2s3p)$ are the dominant channel. The increasing importance of the population of  N$^{3+}(2s3d)$ around 20 keV/u results in a shoulder-like structure in the total electron capture cross sections (see Fig.~\ref{fig1}). Our results for electron capture to N$^{3+}(2s3s)$ and N$^{3+}(2s3d)$ are in good agreement with the experimental data~\cite{McCullough_1995} in the overlapping energies. However, for electron capture to N$^{3+}(2s3p)$, we see important differences between them, with a maximum about 25\% at 1 keV/u. We cannot firmly conclude on that issue, as the experimental data from McCullough \textit{et al.}~\cite{McCullough_1995} were determined by normalizing their total cross sections to previous measurements~\cite{Crandall_1979,Seim_1981,Huq_1989}, among which differences of up to 15\% exist (see Fig.~\ref{fig1}).

We show also in Fig.~\ref{fig2} the spin-averaged $3\ell$-selective electron capture cross sections for N$^{4+}(2p)$ + H($1s$) collisions. It can be observed that the results for N$^{4+}(2p)$ projectile are almost identical with these of N$^{4+}(2s)$ projectile for energies above 4 keV/u. Below, they differ significantly. 

\subsection{Spin-resolved $n\ell$-selective electron capture cross sections}

\begin{figure}[!t]
\begin{center}
\includegraphics[scale=0.5]{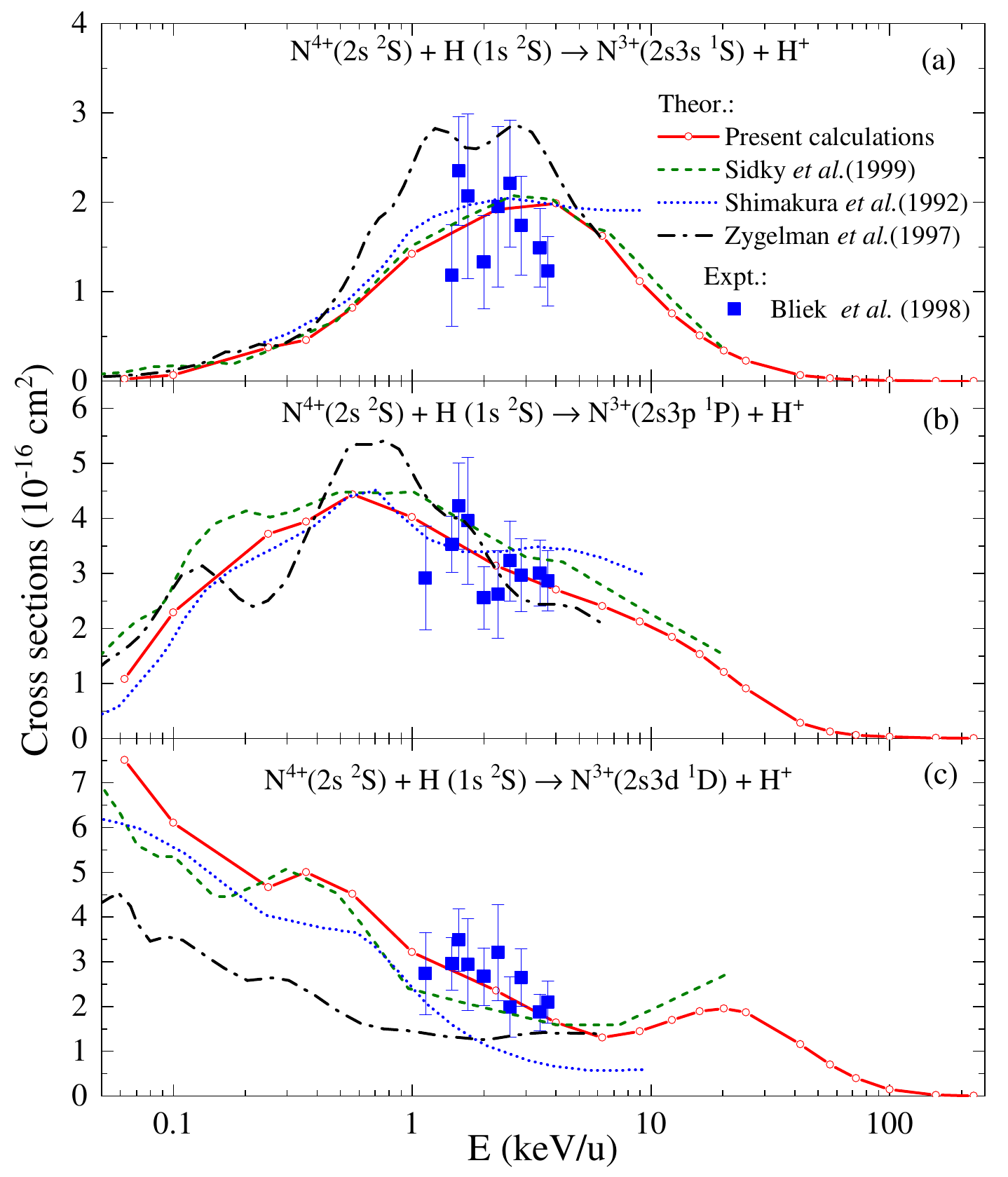}\\
\caption{Cross sections for electron capture to N$^{3+}(2s3\ell$ $^{1}L)$ in N$^{4+}(2s)$ + H($1s$) collisions, the present calculations are plotted as lines with symbols; the MOCC calculations by Zygelman \textit{et al.}~\cite{Zygelman_1997_PRA} and Shimakura \textit{et al.}~\cite{Shimakura_1992} are marked by black dash-dotted and blue short-dotted lines, respectively; the AOCC calculations of Sidky \textit{et al.}~\cite{Sidky_1999} are denoted by green dot-dotted lines. The experimental results of Bliek \textit{et al.}~\cite{Bliek_1998} are shown as symbols and error bars.} \label{fig3}
\end{center}
\end{figure}

Fig.~\ref{fig3} shows the present cross sections for electron capture to the dominant states N$^{3+}(2s3\ell$ $^{1}\rm{L})$ in N$^{4+}(2s)$ + H($1s$) collisions in the case of singlet total spin symmetry, alongside with the available theoretical calculations~\cite{Sidky_1999,Shimakura_1992,Zygelman_1997_PRA} and the only set of existing experimental data~\cite{Bliek_1998}. For electron capture to N$^{3+}(2s3s$ $^{1}\rm{S})$, our cross sections are in good agreement with the experimental data \cite{Bliek_1998} and theoretical calculations~\cite{Sidky_1999,Shimakura_1992} in the respective overlapping energies, except for energies above 4 keV/u, where the MOCC calculations of Shimakura \textit{et al.}~\cite{Shimakura_1992} exceed ours and the AOCC calculations of Sidky \textit{et al.}~\cite{Sidky_1999}. Note that in~\cite{Shimakura_1992}, an approximate electron translation factor  (up to first order in velocity) was employed which can explain the observed discrepancy. For electron capture to N$^{3+}(2s3p$ $^{1}\rm{P})$ and N$^{3+}(2s3d$ $^{1}\rm{D})$, our results show the best overall agreement with the experimental data of Bliek \textit{et al.}~\cite{Bliek_1998}. Some disagreements are seen when comparing the present results to MOCC calculations of Shimakura \textit{et al.}~\cite{Shimakura_1992} and the AOCC calculations of Sidky \textit{et al.}~\cite{Sidky_1999}. The discrepancy is probably due to the insufficient channels included in these two calculations, as discussed above for the total electron capture cross sections. The MOCC calculations by Zygelman \textit{et al.}~\cite{Zygelman_1997_PRA}, where rotational coupling was not included, deviate most from all the other results.

\begin{figure}[!t]
\begin{center}
\includegraphics[scale=0.5]{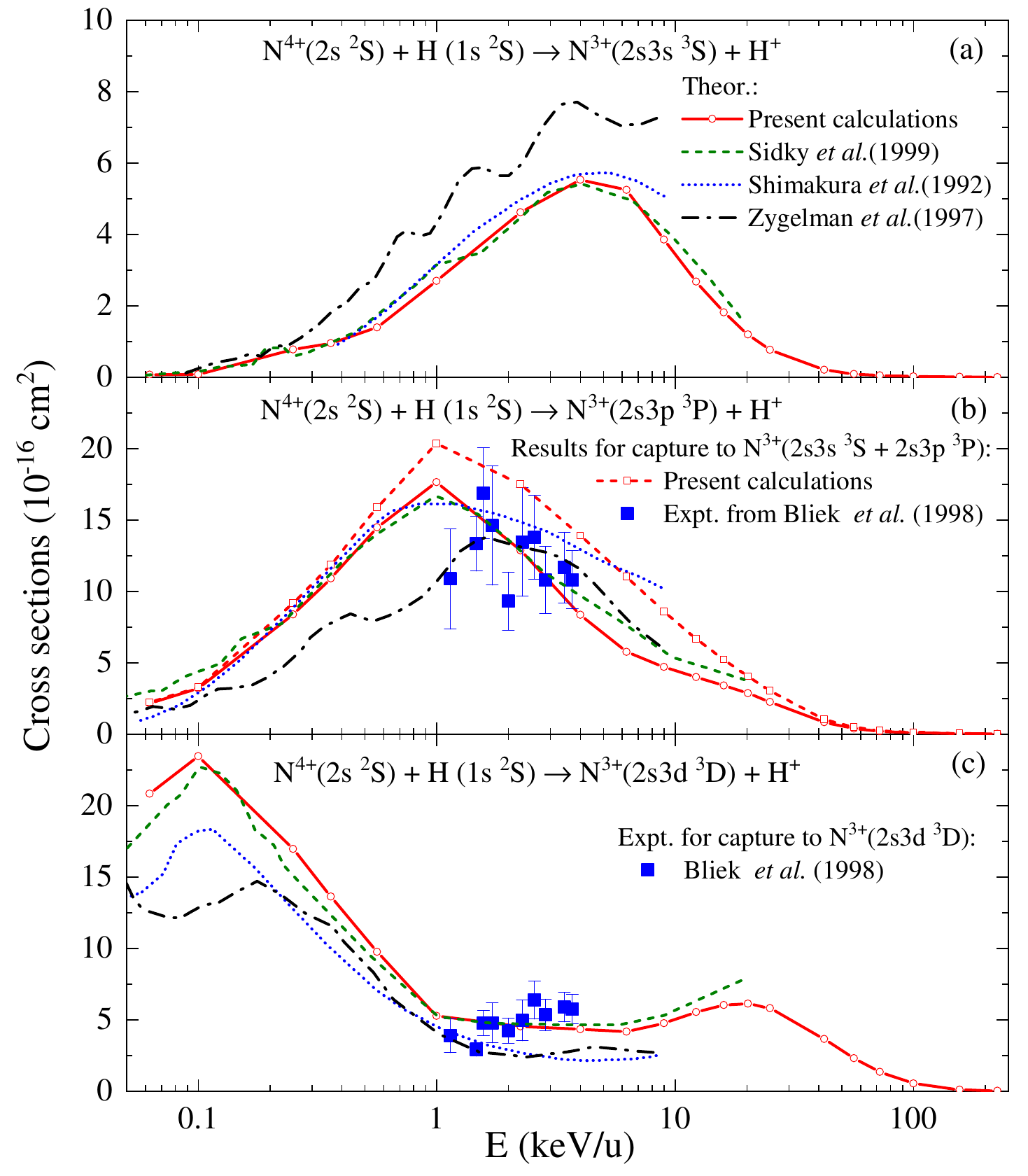}\\
\caption{Same as Fig. \ref{fig3} but for electron capture to N$^{3+}(2s3\ell$ $^{3}L)$ in N$^{4+}(2s)$ + H($1s$). Note that, in (b), the experimental results of Bliek \textit{et al.}~\cite{Bliek_1998} for electron capture to N$^{3+}(2s3s$ $^{3}\rm{S} + 2s3p$ $^{3}\rm{P})$ are shown as blue solid squares, and the corresponding present calculations are plotted as red dashed lines with open squares.}\label{fig4}
\end{center}
\end{figure}

In Fig.~\ref{fig4}, we present our cross sections for electron capture to the dominant states N$^{3+}(2s3\ell$ $^{3}\rm{L})$ in N$^{4+}(2s)$ + H($1s$) collisions in the case of triplet total spin symmetry, together with the available theoretical calculations~\cite{Sidky_1999,Shimakura_1992,Zygelman_1997_PRA} and the only set of existing experimental data~\cite{Bliek_1998}. The comparison between the present results and the other available data is similar to the case of singlet total spin symmetry, except for the electron capture to N$^{3+}(2s3p$ $^{3}\rm{P})$. The measurements by Bliek \textit{et al.}~\cite{Bliek_1998} only provides the summed cross sections for electron capture to N$^{3+}(2s3s$ $^{3}\rm{S})$ and N$^{3+}(2s3p$ $^{3}\rm{P})$, since the measured emission of the $2s3s$ $^{3}$S $\rightarrow$ $2s2p$ $^{3}$P transition results from the capture into both N$^{3+}(2s3s$ $^{3}\rm{S})$ state and N$^{3+}(2s3p$ $^{3}\rm{P})$ state via the cascade $2s3p$ $^{3}$P $\rightarrow$ $2s3s$ $^{3}$S $\rightarrow$ $2s2p$ $^{3}$P. However, the experimental data are smaller than our summed cross sections of N$^{3+}(2s3s$ $^{3}\rm{S})$ and N$^{3+}(2s3p$ $^{3}\rm{P})$, but in good agreement with our present results and those from~\cite{Sidky_1999,Shimakura_1992,Zygelman_1997_PRA} for electron capture to only N$^{3+}(2s3p$ $^{3}\rm{P})$. Without any other data, it is clear that further experimental and theoretical efforts are required to draw definite conclusions on this issue.

\begin{figure}[!h]
\begin{center}
\includegraphics[scale=0.5]{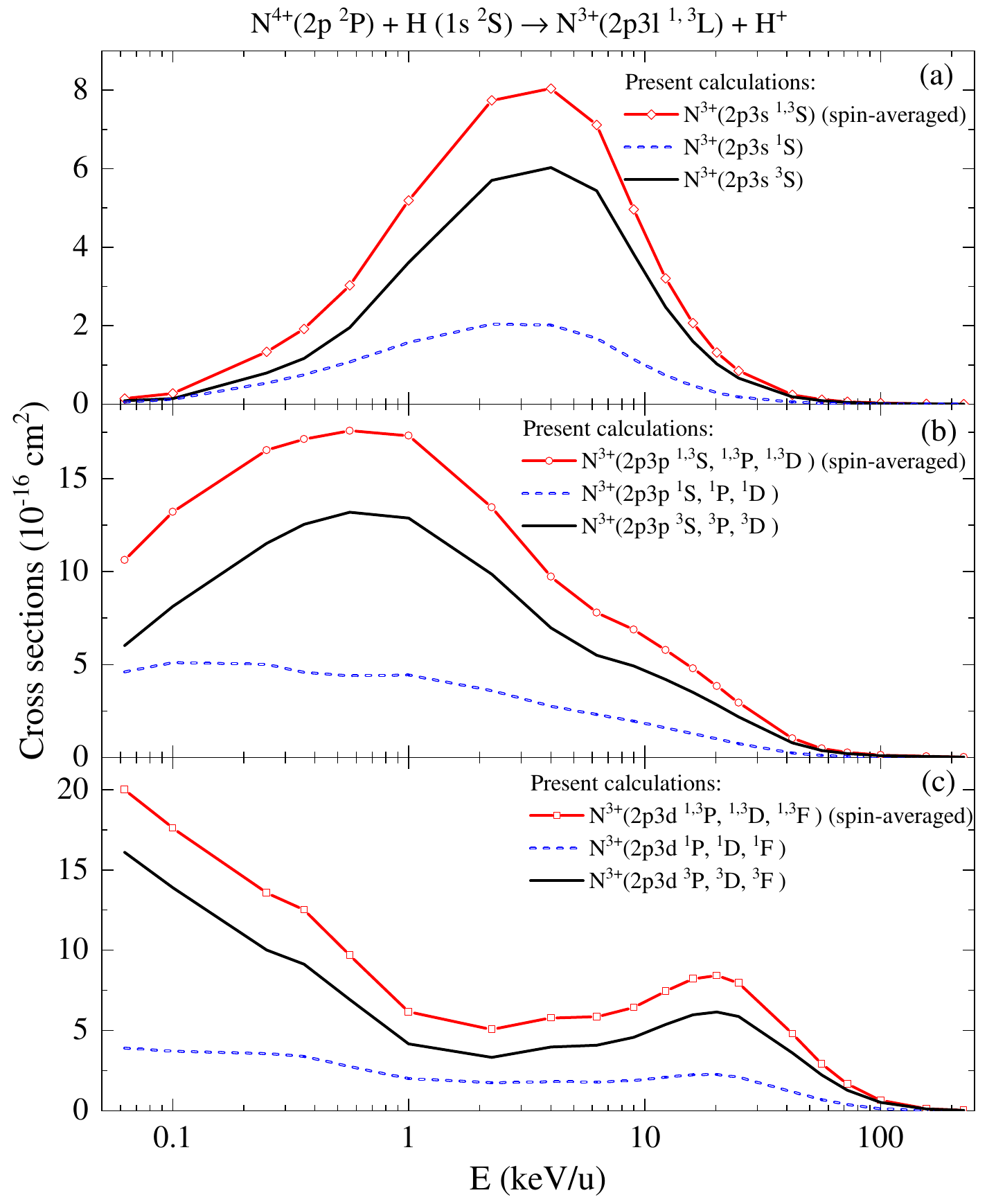}\\
\caption{Cross sections for electron capture to N$^{3+}(2p3\ell$ $^{1,3}L)$ in N$^{4+}(2p)$ + H($1s$) collisions.} \label{fig5}
\end{center}
\end{figure}

We also investigate N$^{4+}(2p)$ + H($1s$) collisions to provide the cross sections for the dominant $3\ell$-selective electron capture processes. These cross sections are reported in Fig.~\ref{fig5}. In the case of triplet spin symmetry, the results for electron capture to N$^{3+}(2p3p$ $^{3}\rm{S}$, $^{3}\rm{P}$, $^{3}\rm{D}$) are dominant in the energy range 0.2 - 10 keV/u, beyond this range, the ones for electron capture to N$^{3+}(2p3d$ $^{3}\rm{P}$, $^{3}\rm{D}$, $^{3}\rm{F}$) are the dominant channel. In the case of singlet spin symmetry, the cross sections for electron capture to N$^{3+}(2p3p$ $^{1}\rm{S}$, $^{1}\rm{P}$, $^{1}\rm{D}$) dominate for energies below 10 keV/u, while the ones for electron capture to N$^{3+}(2p3d$ $^{1}\rm{P}$, $^{1}\rm{D}$, $^{1}\rm{F}$) take over for higher energies. Note that the cross sections for the singlet case are significantly lower than for the tripelet case, as for N$^{4+}(2s)$ + H($1s$) collisions.

\subsection{Target excitation and ionization cross sections}

\begin{figure}[!h]
\begin{center}
\includegraphics[scale=0.5]{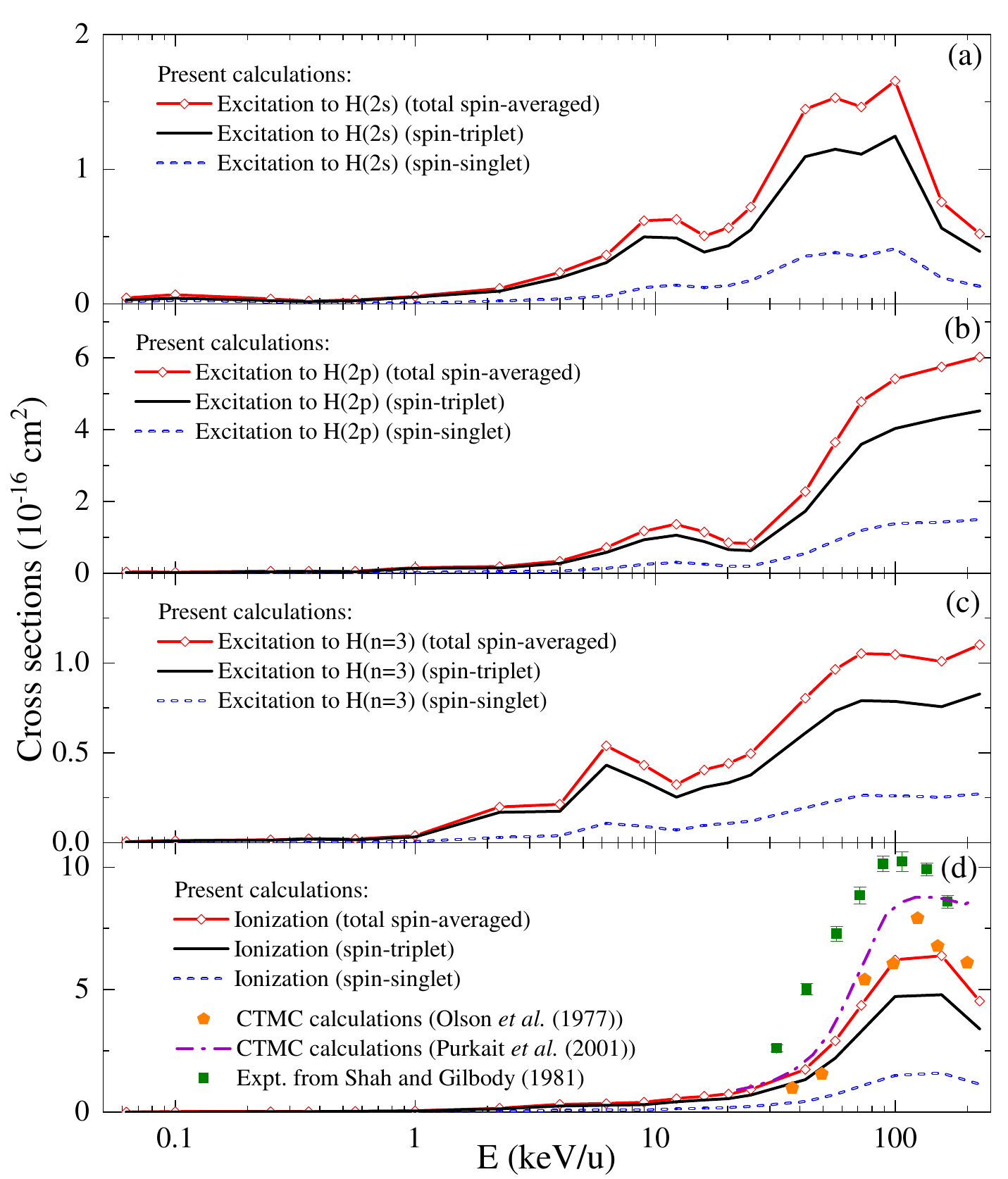}\\
\caption{Spin-resolved and spin-averaged cross sections of N$^{4+}(2s)$ + H($1s$) collisions for target excitation to (a) H($2s$), (b) H($2p$), (c) H($n=3$), and ionization processes in (d). The ionization cross sections from CTMC calculations of Olson \textit{et al.}~\cite{Olson_1977} and Purkait \textit{et al.}~\cite{Purkait_2001_JPB}, as well as the experiment of Shah and Gilbody~\cite{Shah_1981} are also shown in (d).} \label{fig6}
\end{center}
\end{figure}

\begin{figure}[!h]
\begin{center}
\includegraphics[scale=0.5]{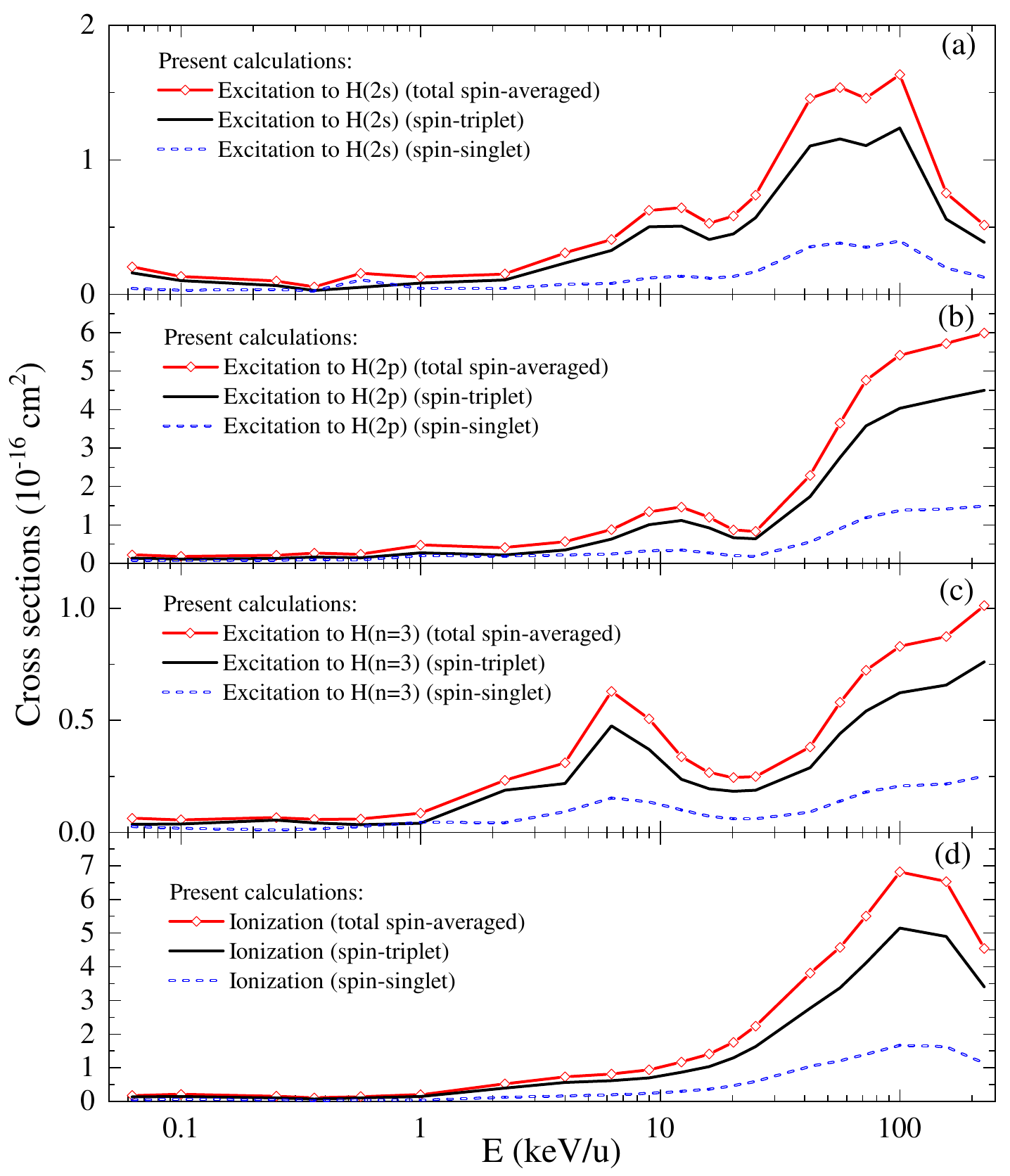}\\
\caption{Same as Fig.~\ref{fig6} but for N$^{4+}(2p)$ + H($1s$) collisions.} \label{fig7}
\end{center}
\end{figure}

We now investigate target excitation of H($1s$) colliding with N$^{4+}(2s)$ and N$^{4+}(2p)$, where no previous theoretical or experimental data was reported. Both spin-resolved and spin-averaged cross sections are presented in Figs.~\ref{fig6}(a-c) for N$^{4+}(2s)$ projectile and Figs.~\ref{fig7}(a-c) for N$^{4+}(2p)$ projectile. As one can see in Fig.~\ref{fig6}, the excitation cross sections are negligibly small for energies below 1 keV/u. Above this energy, H($2p$) production is the strongest target excitation channel, showing the dominant dipolar nature of the interaction with the projectile charge, as the nonrelativistic limit of the so-called Williams-Weizs\"acker virtual photon method~\cite{williams1935,weizsacker1934}. Similar behavior was recently observed in target excitation of He$^{+}$ and H collisions~\cite{Gao_PRA_2024}. In addition, a minimum can be observed in the excitation cross sections around 20 keV/u. This can be assigned to a strong competition between target excitation and electron capture to N$^{3+}(2s3d$ $^{1,3}\rm{D})$, where a maximum can be observed around 20 keV/u (see Fig.~\ref{fig2}). The excitation cross sections for N$^{4+}(2p)$ projectile presented in Fig.~\ref{fig7} show the same trends as those for N$^{4+}(2s)$ projectile, and the same conclusions can be drawn for them.

We next investigate the ionization processes, when the target lost its electron and the projectile remains in the same charge state after the collisions, see Eq.~(\ref{eqc}). Fig.~\ref{fig6}(d) shows the present spin-resolved and spin-averaged cross sections for ionization of atomic hydrogen by N$^{4+}(2s)$ collisions, together with experimental data from Shah and Gilbody~\cite{Shah_1981} and the available CTMC results from Olson \textit{et al.}~\cite{Olson_1977} and Purkait \textit{et al.}~\cite{Purkait_2001_JPB} for comparison. It can be seen that our total spin-averaged cross sections are in good agreement with the CTMC calculations of Purkait \textit{et al.}~\cite{Purkait_2001_JPB} for energies smaller than 60 keV/u and with the ones of Olson \textit{et al.}~\cite{Olson_1977} for energies below 100 keV/u. However, our results lie below the experimental data reported by Shah and Gilbody~\cite{Shah_1981}, by about a factor of 2 at most. The latter discrepancy may be due to the use of GTOs in the present method, which cannot describe the exact boundary conditions of continuum states. However, the experimental data are also larger than the CTMC results. CTMC methods generally describe well ionization processes. Further investigations are needed to resolve this discrepancy. Finally, we present the ionization cross sections for N$^{4+}(2p)$ projectile in Fig.~\ref{fig7}(d) for comparison. These cross sections are similar to those for N$^{4+}(2s)$ projectile. 

We mention here that some two-electron processes are included in the above results: the simultaneous projectile excitation and target ionization, which is included in the total ionization cross sections reported above, contributes to the total ionization cross sections for about 9\% at low impact energies, 15\% for energies in the 1 - 10 keV/u range and 2\% at high impact energies. The two-electron process of projectile excitation and capture is included in the single capture. This process contributes to about 12\% at impact energies less than 0.25 keV/u, 5\% at energies from 0.25 keV/u to 2.25 keV/u and 12\% - 8\% at higher impact energies. Finally, the projectile excitation and target excitation process is included in the target excitation. It represents about 20($\pm$ 7)\% in the energy range 2 - 20 keV/u, 3\% at low impact energies and 1\% at high impact energies of the total target excitation cross sections.

\section{\label{sec4}Conclusions}

Using a two-active-electron semiclassical asymptotic-state close-coupling approach with a configuration interaction treatment to deal with electronic correlation, we have investigated electron capture, target excitation, and ionization processes in ${\rm N}^{4+} (2s)$ and H collisions. In the energy region 0.06 - 225~keV/u, total and state-selective cross sections in both spin-averaged and spin-resolved cases have been reported and compared with available theoretical and experimental data. For electron capture processes, in particular for state-selective cross sections, our results show the best overall agreement with the experimental ones in the different energy ranges available. This demonstrates the importance of a two-electron treatment taking into account electronic correlation and the use of extended basis sets within the close-coupling scheme. 
We also provide new data for target excitation and ionization processes, which are essential to improve our understanding of this relevant collision system. For excitation, H($2p$) production is the strongest target excitation channel, showing the dominant dipolar nature of the interaction with the projectile charge. For the ionization cross sections, our results agree with the experimental data within about a factor of 2. The latter discrepancy may be due to the use of GTOs in the present method, further experimental and theoretical works are, however, needed. Our work provides a complete and consistent set of cross sections over a broad range of collision energies, which can be used for various plasma diagnosis and modeling. 

%These cross sections will be provided in tabulated form on the CollisionDB database~\footnote{The data has been submitted to CollisionDB, see https://db-amdis.org/collisiondb}.

Furthermore, we also present cross sections for electron capture, target excitation, and ionization processes in excited ${\rm N}^{4+} (2p)$ + H collisions. The results exhibit similarities in magnitude and energy dependence with those observed in collisions of ${\rm N}^{4+}(2s)$ projectiles. Since the impurities in the edge plasma environment may be excited due to collisions with electrons and ions/atoms, the reported cross sections for ${\rm N}^{4+} (2p)$ projectile should be also of relevance in the studies of properties of fusion plasma.

\section{\label{sec5}ACKNOWLEDGMENTS}
This work is supported by the National Natural Science Foundation of China (Grants No.11934004 and No. 12374229), and the startup project from Hangzhou Normal University. C.C.J. is supported by a China Scholarship Council (CSC) scholarship within a CSC-SU program.

\bibliography{ref} 
% Produces the bibliography via BibTeX.

%\clearpage
\end{document}